\documentclass[preprint,prc,aps,showpacs,showkeys]{revtex4}
\usepackage{epsfig}

\begin{document}

\title{Hadron Masses in Medium and Neutron Star Properties}

\author{
C. Y. Ryu$^a$, 
C. H. Hyun$^{a,b}$\footnote{e-mail:hch@meson.skku.ac.kr}, 
S. W. Hong$^a$,
B. K. Jennings$^c$}
\affiliation{
(a) Department of Physics and Institute of Basic Science, 
Sungkyunkwan University, Suwon 440-746, Korea, \\
(b) School of Physics, Seoul Nat'l University,
Seoul 151-742, Korea \\
(c) TRIUMF, 4004 Wesbrook Mall, Vancouver,
British Columbia, Canada V6T 2A3}

\date{February 3, 2005}

\begin{abstract}
We investigate the properties of the neutron
star with relativistic mean field models.
We incorporate 
in the quantum hadrodynamics and in the quark-meson coupling models
a possible reduction of meson masses in nuclear matter.
The equation of state for neutron star matter is 
obtained and is employed in Oppenheimer-Volkov equation
to extract the maximum mass of the stable neutron star.
We find that the equation of state, the composition 
and the properties of the neutron stars
are sensitive to the values of the meson masses in medium.
\end{abstract}
\maketitle
\setcounter{footnote}{0}

\section{Introduction}

The state of matter at extremely hot or dense conditions is
one of the most fundamental questions in physics.
Terrestrial nuclei of heavy elements provide a glimpse at such
extreme states, but there are far-more-extreme states of matter 
in our universe.
A neutron star, which may be regarded as
a huge nucleus, is one example of such an
extreme state of matter.
The state of matter can be characterized by 
the equation of state (EoS).
From the microscopic point of view the EoS is determined by how the
constituent particles of the matter interact with each other. 
Therefore the state of matter at extreme conditions can be understood
when we have enough knowledge about the properties of the
constituent particles and how they interact in such conditions.

In the early 90's, Brown and Rho 
proclaimed the scaling property of hadron masses in 
dense medium in terms of the scale invariance of the effective 
lagrangian \cite{brprl91}.
They showed that the ratios of the in-medium masses of the nucleon, 
$\sigma$-, $\omega$- and $\rho$-mesons to their masses in free
space are
approximately equal to each other at around the nuclear saturation 
density.
Afterwards, changes in the hadron masses in medium were calculated
in the framework of QCD-sum rule \cite{hlprc92} 
and quark-meson-coupling (QMC) models \cite{sttprc97}.
In Refs.~\cite{hlprc92} and \cite{sttprc97}, masses of $\rho$ and $\omega$
mesons were calculated at the nuclear saturation density,
and similar amount of 
mass reduction from the free mass, $m^*_{\rho, \omega}/m_{\rho, \omega} 
\simeq 0.8$ was predicted.
Experimentally, microscopic 
properties of hadrons as well as the states of matter at 
hot or dense environment can be probed in 
relativistic heavy-ion collisions, and
possible meson mass reductions in nuclear matter 
were reported 
through the dileptonic decay of $\rho$ and $\omega$ mesons in 
the CERES/NA45 \cite{ceres} and KEK-PS E325 \cite{kek}
experiments.
Also, recent astronomical observations of 
binary systems yield the mass of compact objects as
$M_{\rm J0751+1807} = (2.2 \pm 0.2) M_\odot$ \cite{stairs04}
and $M_{\rm 4U1700-37} = (2.44 \pm 0.27) M_\odot$ \cite{quain-aa03},
of which the former is believed to be a neutron star.
Such a large mass of the neutron star requires a substantially
stiff EoS.

Motivated by these theoretical and experimental studies,
in this work we take into account the reduction of meson
masses in the nuclear models such as quantum hadrodynamics (QHD)
and QMC, and apply these models to the neutron star matter.
The nuclear models are calibrated to the bulk properties of nuclear 
matter at the saturation density, such as 
saturation density $\rho_0$ (= 0.17 fm$^3$), 
binding energy $E_B$ (= 16.0 MeV), symmetry energy $a_{\rm sym}$
(= 32.5 MeV) and compression modulus $K$ (200 $\sim$ 300 MeV).
We take into account the effect of meson mass changes in the following ways.
First, we use the meson masses in free space for the nuclear models 
considered here as is usually done in most works. 
Secondly, we adopt the Brown-Rho (BR) scaling law \cite{brprl91},
\begin{eqnarray}
\frac{m^*_N}{m_N} \simeq \frac{m^*_\sigma}{m_\sigma} \simeq
\frac{m^*_\omega}{m_\omega} \simeq \frac{m^*_\rho}{m_\rho}
\label{eq:br}
\end{eqnarray}
and assume a density-dependent scaling function 
for the meson masses \cite{sbmr97}.
The nuclear models that employ the scaling function are denoted by the 
abbreviation ``S". 
Thirdly, in using the QMC type models we may
treat the heavy mesons ($\rho$ and $\omega$) as meson bags 
(denoted by ``MB") composed of
a quark and an anti-quark \cite{stprc95}.
In such cases, heavy mesons
are treated in consistent with the description of 
the nucleon in the framework of the QMC.
After calibrating different nuclear models for nuclear matter 
we apply them to the neutron star matter 
and see how the reduction of meson masses in matter
may influence the EoS of the neutron star matter 
and the properties of the neutron star.

In Sec. 2, we briefly describe the nuclear models 
considered in this work and fix the parameters of the models and 
show the resultant properties of the nuclear
matter at the saturation density.
In Sec. 3, we apply the various nuclear models fixed in Sec. 2
to the neutron star matter and discuss the results and 
implications for the properties of the neutron star.
A summary is given in Sec. 4.

\section{Models for Nuclear Matter} 

In this work we will consider 5 different models.
We refer to them as QHD, QHD-S; MQMC, MQMC-S and MQMC-MB.
All the models have three coupling constants 
$g_{\sigma N}$, $g_{\omega N}$, and $g_{\rho N}$ 
(or $g^q_\sigma$, $g^q_\omega$, and $g^q_\rho$) 
for the interaction of nucleons (or quarks)
with $\sigma$-, $\omega$- and $\rho$-mesons, respectively.
$\sigma$- and $\omega$-coupling constants are fitted to 
reproduce the given values of $\rho_0$
and $E_B$, and the coupling to $\rho$-meson is determined by
$a_{\rm sym}$. 
Details of each model are explained in the following subsections.

\subsection{QHD}

The lagrangian for QHD in the mean field approximation reads
\begin{eqnarray}
L^{\rm MF}_{\rm QHD}
&=& \bar{\psi}_N [i \gamma^\mu \partial_\mu - ( m_N
- g_{\sigma N}\, \bar{\sigma}) - g_{\omega N}\,
\gamma^0\, \bar{\omega}_0
- \frac{1}{2}\, g_{\rho N}\, \gamma^0\, \bar{b}_{03}\, \tau_3]
\psi_N \nonumber \\
&& - \frac{1}{2} m^2_\sigma\, \bar{\sigma}^2
- \frac{1}{3} m_N\, b\, (g_{\sigma N}\, \bar{\sigma})^3
- \frac{1}{4} c\, (g_{\sigma N}\, \bar{\sigma})^4 \nonumber \\ &&
+ \frac{1}{2} m^2_\omega\, \bar{\omega}^2_0
+ \frac{1}{2} m^2_\rho\, \bar{b}^2_{03}.
\label{eq:qhdlagrangian}
\end{eqnarray}
The masses $m_N$, $m_\sigma$, $m_\omega$ and $m_\rho$
refer to the values in free space, 939, 550, 783 and 
770 MeV, respectively.
The effective mass of the nucleon is defined as
\begin{eqnarray}
m^*_N({\rm QHD}) \equiv m_N - g_{\sigma N}\, \bar{\sigma}.
\label{eq:meff-qhd}
\end{eqnarray}
The meson fields are determined from the 
equations of motion in the mean field approximation ;
\begin{eqnarray}
\bar{\sigma} &=& 
\frac{g_{\sigma N}}{m^2_\sigma}\, \rho_s
- \frac{m_N}{m^2_\sigma}\, b\, g^3_{\sigma N}\, \bar{\sigma}^2
- \frac{1}{m^2_\sigma}\, c\, g^4_{\sigma N}\, \bar{\sigma}^3,
\label{eq:qhdscc} \\
\bar{\omega}_0 &=& \frac{g_{\omega N}}{m^2_\omega}\, 
\frac{1}{3 \pi^2}
\sum_{N=n,p} k^3_N = \frac{g_{\omega N}}{m^2_\omega}\, \rho, 
\label{eq:omega} \\
\bar{b}_{03} &=& \frac{g_{\rho N}}{m^2_\rho}\,
\frac{1}{3 \pi^2}
\sum_{N=n,p} I_{N3} k^3_N
= \frac{1}{2} \frac{g_{\rho N}}{m^2_\rho}\,
(\rho_p - \rho_n),
\label{eq:rho}
\end{eqnarray}
where $I_{N3}$ is 1/2 (-1/2) for the proton (neutron).
\begin{eqnarray}
\rho_s = \frac{1}{\pi^2} \sum_{N=n,p}
\int^{k_N}_0 \frac{m^*_N}{\sqrt{k^2 + m^{*2}_N}}\, k^2\, dk
\label{eq:scalardensity}
\end{eqnarray} 
is the scalar density and $k_N$ is the Fermi momentum of the 
nucleon at a given density.
Cubic and quartic self-interaction terms of the $\sigma$-meson 
are included in Eq. (\ref{eq:qhdlagrangian})
to produce reasonable $K$ and $m^*_N$ values at 
the saturation density.
The coupling constants fixed to produce the saturation properties
and the resulting $m^*_N$ and $K$ are listed in Tab. \ref{tab:qhd}.

\subsection{QHD-S}

We now consider the meson mass changes in matter and 
adopt the model proposed in Ref.~\cite{sbmr97}  where 
the BR-scaling law is incorporated in the original
QHD.
The model lagrangian reads
\begin{eqnarray}
L^{\rm MF}_{\rm QHD-S} &=& \bar{\psi}_N [
i \gamma^\mu \partial_\mu - (M^*_N - g_{\sigma N}\, \bar{\sigma})
- g^*_{\omega N}\,  \gamma^0\,  \bar{\omega}_0
- \frac{1}{2} g_{\rho N}\,  \gamma^0\,  \bar{b}_{03}\,  \tau_3]
\psi_N \nonumber \\
&& - \frac{1}{2} m^{*2}_\sigma\,  \bar{\sigma}^2
+ \frac{1}{2} m^{* 2}_{\omega}\,  \bar{\omega}^2_0
+ \frac{1}{2} m^{* 2}_\rho\,  \bar{b}^2_{03}.
\end{eqnarray}
The BR-scaling law is parametrized as \cite{sbmr97}
\begin{eqnarray}
\frac{M^*_N}{m_N} = \frac{m^*_\sigma}{m_\sigma}
= \frac{m^*_\omega}{m_\omega}
= \frac{m^*_\rho}{m_\rho}
= \left( 1 + y \frac{\rho}{\rho_0}\right)^{-1}.
\label{eq:brscaling}
\end{eqnarray}
The effective mass of the nucleon is defined as 
\begin{eqnarray}
m^*_N(\mbox{QHD-S}) = M^*_N - g_{\sigma N}\, \bar{\sigma}.
\label{eq:efmnqhdbr}
\end{eqnarray}
(Note the use of $M^*_N$ in place of $m_N$ of Eq.~(\ref{eq:meff-qhd}).)
Coupling constants may change with nuclear matter density.
Such density dependences of coupling constants as well as masses
have been considered in previous \cite{glen-prc92,fuchs-prc95,typel-npa99}
and very recent \cite{liu-04,kolo-04} works
within the framework of relativistic 
mean field theory.
In Ref.~\cite{sbmr97} , $g_{\omega N}$ is assumed to vary
with density to satisfy $g^*_{\omega N}/g_{\omega N}
\simeq m^*_\omega/m_\omega$ at around the saturation
density. 
$g^*_{\omega N}$ scaling is expressed as \cite{sbmr97} 
\begin{eqnarray}
\frac{g^*_{\omega N}}{g_{\omega N}} = \left( 1 + z \frac{\rho}{\rho_0}
\right)^{-1}.
\label{eq:gscale}
\end{eqnarray}

The parameters for the QHD-S model
and the resulting properties at the saturation are summarized 
in Tab. \ref{tab:qhd}. 
One can see that both QHD and QHD-S models give $K$ values 
in the acceptable range.
$m^*_M/m_M$ turns out to be 0.781, which is close to the values
obtained by previous studies \cite{hlprc92,sttprc97}.
$m^*_N/m_N$ from QHD-S is smaller than that from QHD because of
the scaling in Eq. (\ref{eq:brscaling}).
\begin{table}
\begin{tabular}{|c|c|c|c|c|c|c|c||c|c|c|}\hline
Model & $g_{\sigma N}$ & $g_{\omega N}$ & $g_{\rho N}$
& $b \times 100$ & $c \times 100$ & $y$ & $z$ 
& $m^*_N/m_N$ & $m^*_M/m_M$ & $K$ \\
\hline\hline
QHD  & 8.11 & 8.36  & 7.85 & 0.3478 & 1.328 & $\cdot$ & 
$\cdot$ & 0.773 & 1.0 & 310.8 \\ \hline
QHD-S  & 5.30 & 15.30 & 7.52 & $\cdot$ & $\cdot$ & 
0.28 & 0.31 & 0.667 & 0.781 & 264.5 \\ \hline
\end{tabular}
\caption{Parameters and the saturation properties for the 
QHD and QHD-S models. The subscript $M$ for $m^*_M/m_M$ refers to 
$\sigma$, $\omega$ and $\rho$ mesons.}
\label{tab:qhd}
\end{table}

\subsection{MQMC}

In the QMC model \cite{guichon88}, 
interactions between the nucleons (or bags) are
mediated by the exchange of mesons which couple with the quarks 
in the nucleon bags.
The mean field lagrangian for non-overlapping spherical 
bags in dense matter may be written as
\begin{eqnarray}
L^{\rm MF}_{\rm QMC} &=& \bar{\psi}_q
[ i \gamma^\mu \partial_\mu - (m^0_q - g^q_\sigma\, \bar{\sigma})
- g^q_\omega\, \gamma^0\, \bar{\omega}_0
- \frac{1}{2}g^q_\rho\, \gamma^0\, \bar{b}_{03}\, \tau_3 - B]  \times
\theta_V(R - r) \psi_q \nonumber \\ & &
 - \frac{1}{2} m^2_\sigma\, \bar{\sigma}^2
+ \frac{1}{2} m^2_\omega\, \bar{\omega}^2_0
+ \frac{1}{2} m^2_\rho\, \bar{b}^2_{03},
\label{eq:qmclagrangian}
\end{eqnarray}
where $g^q_M$ ($q = u,\, d\, ; M = \sigma,\, \omega,\, \rho$)
is the coupling constant for the quark-meson interaction.
$B$ is the bag constant and $R$ is the bag radius within which
quarks are confined.
The effective mass of the nucleon in the QMC is given by
\begin{eqnarray}
m^*_N({\rm QMC}) &=& \sqrt{\left( E^N_{bag} \right)^2
- \sum_q \frac{x^2_q}{R^2}}, 
\label{eq:qmcmstar}\\
E^N_{bag} &=& \sum_q \frac{\Omega_q}{R} -
\frac{Z_N}{R} + \frac{4}{3}\pi R^3 B, \\
\Omega_q &=& \sqrt{x^2_q + R^2 m^{*2}_q},\ \ \
(m^*_q = m^0_q - g^q_\sigma \bar{\sigma})
\label{eq:Omega-q}.
\end{eqnarray}
$x_q$ is the eigen energy of the quarks in the bag determined by
the boundary conditions at $r = R$. 
$Z_N$ is a phenomenological constant that incorporates
the effects not explicitly taken into account, including 
zero-point motion.

The scalar and the vector potentials obtained from the QMC model
were much smaller than those from the QHD,
which would result in too weak a spin-orbit potential to explain
the spin-orbit splittings in finite nuclei and 
spin observables in nucleon-nucleus scattering.
To circumvent these shortcomings, density dependent bag constants
were introduced \cite{jjprc96}. 
The QMC model with a density dependent bag constant is called the 
modified QMC (MQMC) model.
We employ the direct coupling model of Ref.~\cite{jjprc96} where 
the bag constant reads
\begin{eqnarray}
B = B_0 \left( 1 - g^B_\sigma \frac{4}{\delta}
\frac{\bar{\sigma}}{m_N} \right)^\delta.
\label{eq:bagconstant}
\end{eqnarray}
$\bar{\sigma}$ value is determined from the 
self-consistency condition (SCC)
\begin{eqnarray}
\bar{\sigma} =
3\, \frac{g^q_\sigma}{m^2_\sigma} \, \rho_s
\left[
C_N(\bar{\sigma}) +
\frac{g^B_\sigma}{g^q_\sigma} \frac{E^N_{bag}}{m^*_N}
\frac{16\pi}{9} R^3 \frac{B}{m_N}
\left( 1 - \frac{4}{\delta}
\frac{g^B_\sigma \bar{\sigma}}{m_N} \right)^{-1}
\right]
\label{eq:sccmqmc}
\end{eqnarray}
with
\begin{eqnarray}
C_N(\bar{\sigma}) = \frac{E^N_{bag}}{m^*_N}
\left[\left( 1 - \frac{\Omega_q}{E^N_{bag} R}\right)
S(\bar{\sigma}) + \frac{m^*_q}{E^N_{bag}}\right]
\label{eq:scc2}
\end{eqnarray}
and
\begin{eqnarray}
S(\bar{\sigma}) = \frac{\Omega_q/2 + R\, m^*_q\,
(\Omega_q - 1)}{\Omega_q\, (\Omega_q - 1) 
+ R\, m^*_q/2}.
\label{eq:scc3}
\end{eqnarray}
$B_0$ and $Z_N$ are fitted to reproduce the free 
nucleon mass with the stability condition
\begin{eqnarray}
\left. \frac{\partial\, m^*_N}{\partial \, R} 
\right|_{R = R_0} = 0
\label{eq:minimum}
\end{eqnarray}
where $m^*_N$ is evaluated by Eq.~(\ref{eq:qmcmstar}).
Choosing $R_0 = 0.6$ fm, we obtain
$B_0^{1/4} = 188.1$ MeV and $Z_N = 2.030$.
$g^B_\sigma$ and $\delta$, $g^q_\sigma$ and 
$g^q_\omega$ are adjusted to produce saturation properties and
reasonable values of $K$ and $m^*_N$.
$g^q_\rho$ is fitted for the symmetry energy to be reproduced.

\subsection{MQMC-S}

As a simple way to incorporate the reduction of meson mass 
in medium, we assume
\begin{eqnarray}
\frac{m^*_\sigma}{m_\sigma} =
\frac{m^*_\omega}{m_\omega} =
\frac{m^*_\rho}{m_\rho} =
\left(1 + y \frac{\rho}{\rho_0} \right)^{-1}.
\label{eq:mscaling}
\end{eqnarray}
The scaling parameter $y$ is determined to satisfy
Eq.~(\ref{eq:br}) at around the saturation density
with $m^*_N$ given by Eq.~(\ref{eq:qmcmstar}).
The lagrangian for this model can be easily obtained by replacing
the meson masses in Eq.~(\ref{eq:qmclagrangian})
with the medium-modified values in Eq.~(\ref{eq:mscaling}).
Since the SCC is obtained from the minimization of 
the energy density with respect to the variation of $\bar{\sigma}$,
the SCC of the MQMC-S is of the same form as that of MQMC,
but with $m_\sigma$ replaced by $m^*_\sigma$ given by 
Eq.~(\ref{eq:mscaling}).
The model parameters fixed to produce 
saturation properties are listed in Tab. \ref{tab:mqmc}.
\subsection{MQMC-MB}

In the framework of MQMC, we may treat
vector mesons as MIT bags composed of a quark and an anti-quark. 
We refer to such a model as MQMC-MB.
The parameters $B_0$, $R_0$ and $Z_M$ ($M = \rho,\ \omega$)
for the meson bags can differ depending on the mesons, but
for simplicity we fix $B_0$ and $R_0$ as those values for the nucleon 
and treat only $Z_M$ as a parameter for each meson \cite{stprc95}. 
The effective mass of each meson can be written as 
\begin{eqnarray}
m^*_M &=& 
\sqrt{\left(E^M_{bag}\right)^2 - 2 \frac{x_q^2}{R^2}},
\label{eq:mesonmass} \\
E^M_{bag} &=& 2 \frac{\Omega_q}{R} 
- \frac{Z_M}{R} + 
\frac{4}{3}\pi\, R^3\, B.
\label{eq:mesonbag}
\end{eqnarray}
$Z_M$ is fixed so that $m^*_M$ defined 
as in Eq.~(\ref{eq:mesonmass}) recovers the mass of the $\omega$- 
and $\rho$-mesons in free space. 
$Z_\omega = 0.7904$ and $Z_\rho = 0.8154$ are obtained.
The lagrangian for the model can be written by replacing the 
meson masses in Eq.~(\ref{eq:qmclagrangian}) 
with $m^*_M$ in Eq.~(\ref{eq:mesonmass}).
In the MQMC-MB meson masses are functions of $\bar{\sigma}$,
which modifies the SCC as
\begin{eqnarray}
\bar{\sigma} &=&
3\, \frac{g^q_\sigma}{m^2_\sigma} \, \rho_s
\left[
C_N(\bar{\sigma}) +
\frac{g^B_\sigma}{g^q_\sigma} \frac{E^N_{bag}}{m^*_N}
\frac{16\pi}{9} R^3 \frac{B}{m_N}
\left( 1 - \frac{4}{\delta}
\frac{g^B_\sigma \bar{\sigma}}{m_N} \right)^{-1}
\right] \nonumber \\
&-& 18\, \frac{g^q_\sigma}{m^2_\sigma}
\frac{g^{q 2}_\omega}{m^{* 3}_\omega}\, \rho^2
\left[ C_\omega(\bar{\sigma}) +
\frac{g^B_\sigma}{g^q_\sigma} \frac{E^\omega_{bag}}{m^*_\omega}
\frac{8\pi}{3} R^3 \frac{B}{m_N}
\left( 1 - \frac{4}{\delta}
\frac{g^B_\sigma \bar{\sigma}}{m_N} \right)^{-1}
\right] \nonumber \\
&-& 2\, \frac{g^q_\sigma}{m^2_\sigma}
\frac{g^{q 2}_\rho}{m^{* 3}_\rho}\, \left(
\sum_{N=n,p} I_{N3} \rho_N \right)^2
\left[ C_\rho(\bar{\sigma}) +
\frac{g^B_\sigma}{g^q_\sigma} \frac{E^\rho_{bag}}{m^*_\rho}
\frac{8\pi}{3} R^3 \frac{B}{m_N}
\left( 1 - \frac{4}{\delta}
\frac{g^B_\sigma \bar{\sigma}}{m_N} \right)^{-1}
\right]
\end{eqnarray}
where
\begin{eqnarray}
C_M(\bar{\sigma}) \equiv
\frac{E^M_{bag}}{m^*_M}
\left[ \left( 1 - \frac{\Omega_q}{E^M_{bag}\, R}\right)
S(\bar{\sigma}) + \frac{m^*_q}{E^M_{bag}} \right]. 
\end{eqnarray}

\begin{table}[tbp]
\begin{tabular}{|c|c|c|c|c|c||c|c|c|}\hline
Model & $g^q_\sigma$ & $g^q_\omega$ & $g^B_\sigma$ &
$g^q_\rho$ & $y$ & $m^*_N/m_N$ & $m^*_M/m_M$ & $K$ \\ \hline
MQMC & 1.0 & 2.71 & 6.81 & 7.89 & $\cdot$ & 0.783 & 1.0 & 285.5 
\\ \hline
MQMC-S & 1.0 & 2.31 & 5.51 & 6.09 & 0.28 &
0.758 & 0.781 & 591.5
\\ \hline
MQMC-MB & 1.0 & 1.77 & 5.44 & 8.15 & $\cdot$ & 0.852 & 0.861 & 324.1
\\ \hline
\end{tabular}
\caption{
The parameters and the saturation properties from MQMC-type models.
The subscript $M$ denotes $\sigma$, $\omega$ and 
$\rho$ mesons for MQMC-S.
For MQMC-MB, $M$ refers to $\omega$ and $\rho$ mesons.
$B^{1/4}_0 = 188.1$ MeV and $Z_N = 2.030$ are used for all three 
cases.}
\label{tab:mqmc}
\end{table}

The parameters that can produce the saturation at $\rho_0$ are 
listed in Tab. \ref{tab:mqmc}.
MQMC and MQMC-MB give us acceptable results for $K$ and $m^*_N$,
but the $K$ value for MQMC-S is as large as twice of 
usually accepted ranges of values.
To get a value of $K$ within a reasonable range,
we have tried many numerical searches over a wide range of parameter space,
but no solution has been found so far.

\section{Application to the Neutron Star}

We now apply the five nuclear models
described in Sect. II to the neutron star matter.
Two basic assumptions of the neutron star matter are 
the charge neutrality and the $\beta-$equilibrium.
If we assume that only the nucleons and light leptons
exist in the neutron star, charge neutrality is expressed as
\begin{eqnarray}
\rho_p = \sum_{l = e, \mu} \rho_l,
\label{eq:charge0}
\end{eqnarray}
where $\rho_i$ is the number density of particle 
$i\, (=p,\, e,\, \mu)$.
Under $\beta-$equilibrium, the processes
\begin{eqnarray}
n \rightarrow p + e^- + \bar{\nu}_e \,\,\,{\rm and}
\,\,\,
p + e^- \rightarrow n + \nu_e
\label{eq:durca}
\end{eqnarray}
occur at the same rate.
The condition can be satisfied when the chemical potentials before 
and after the decay are the same.
The chemical potential of each particle reads
\begin{eqnarray}
\mu_n &=& \sqrt{k^2_n + m^{*2}_N}
+ g_{\omega N}\, \bar{\omega}_0
- \frac{1}{2}\, g_{\rho N}\, \bar{b}_{03}, \\
\mu_p &=& \sqrt{k^2_p + m^{*2}_N}
+ g_{\omega N}\, \bar{\omega}_0
+ \frac{1}{2}\, g_{\rho N}\, \bar{b}_{03}, \\
\mu_l &=& \sqrt{k^2_l + m^2_l},
\end{eqnarray}
where
$k_l$ is the Fermi momentum of the lepton
$l$ ($= e,\ \mu$).
The chemical equilibrium condition is expressed as
\begin{eqnarray}
\mu_n = \mu_p + \sum_l \mu_l.
\label{eq:chemeq}
\end{eqnarray}

\subsection{EoS \label{section:eos}}
\begin{figure}[tbp]
\begin{center}
\epsfig{file=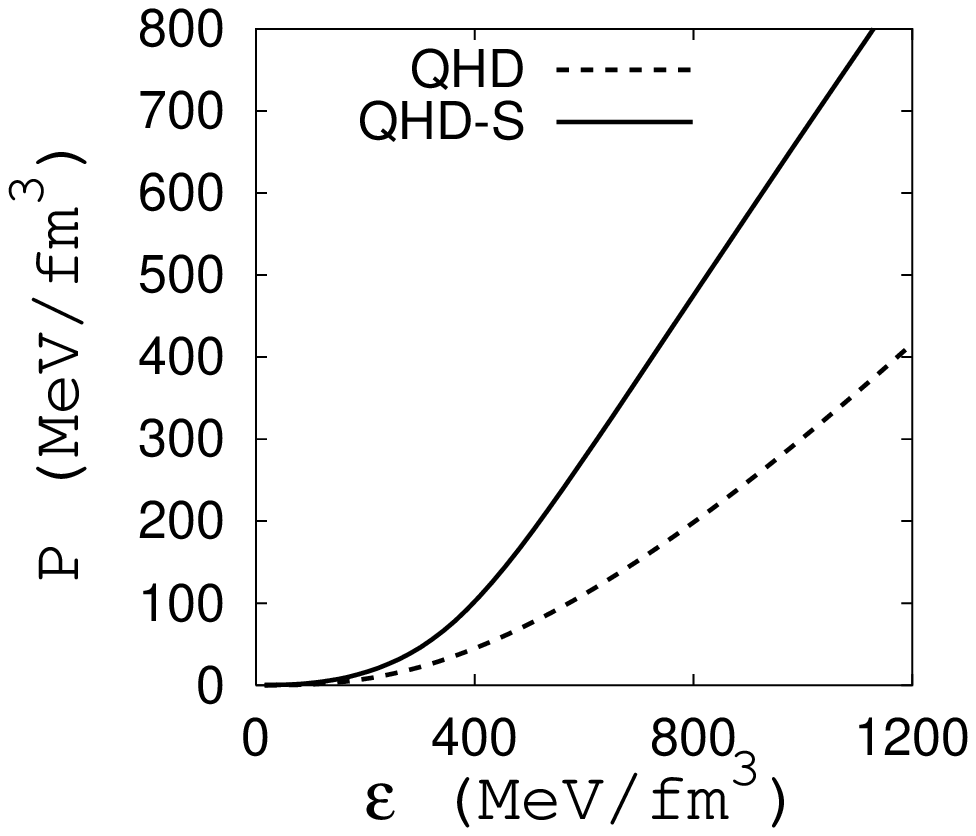,width=8.0cm}
\epsfig{file=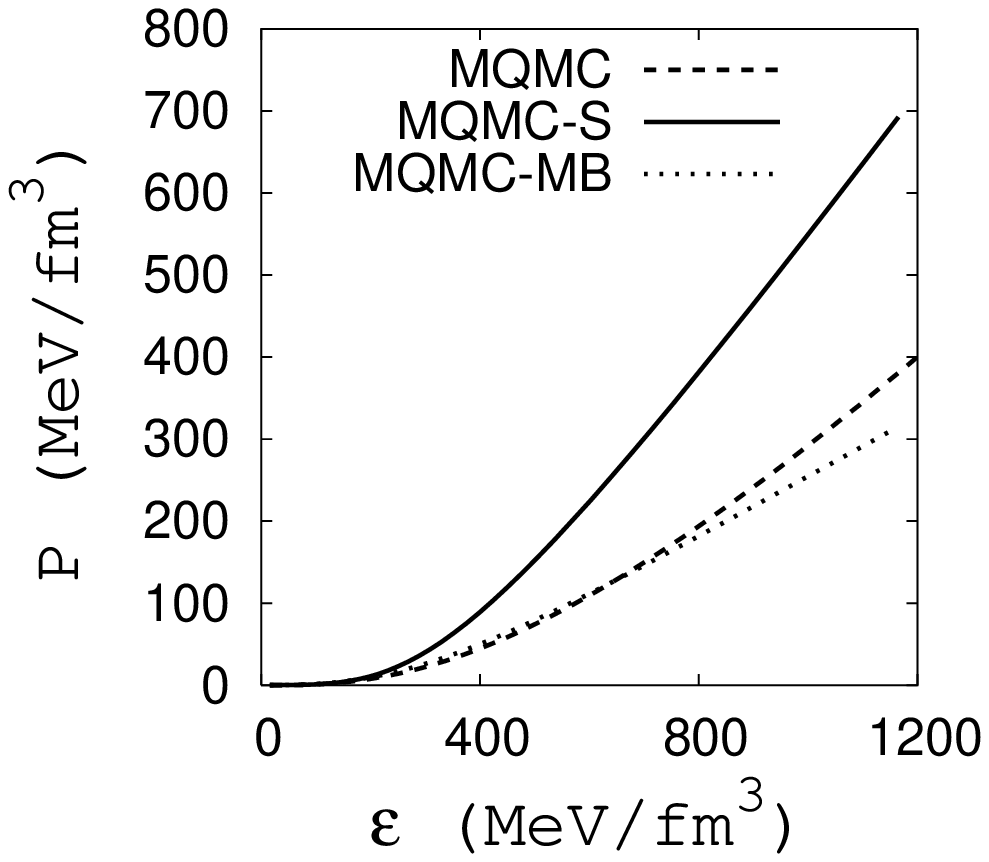,width=8.0cm}
\caption{EoS from five different models. 
\label{fig:eos}}
\end{center}
\end{figure}

The EoS tells us the pressure ($P$) as a function of 
energy density ($\varepsilon$),
which are the diagonal elements of the energy-momentum tensor.
The energy density and the pressure in the QHD model read
\begin{eqnarray}
\varepsilon &=& \frac{1}{2} m^2_\sigma\, \bar{\sigma}^2
+ \frac{1}{3} m_N\, b\, (g_{\sigma N}\, \bar{\sigma})^3
+ \frac{1}{4} c\, (g_{\sigma N}\, \bar{\sigma})^4
+ \frac{1}{2} m^2_\omega\, \bar{\omega}^2_0
+ \frac{1}{2} m^2_\rho\, \bar{b}^2_{03} \nonumber \\
&+& \frac{1}{\pi^2}
\sum_{N = n, p} \int^{k_N}_0 \sqrt{k^2 + m^{*2}_N}\ k^2 dk
+ \frac{1}{\pi^2}
\sum_l \int^{k_l}_0 \sqrt{k^2 + m^2_l}\ k^2 dk,
\label{eq:eneqhd}
\end{eqnarray}
\begin{eqnarray}
P &=&
- \frac{1}{2} m^2_\sigma\, \bar{\sigma}^2
- \frac{1}{3} m_N\, b\, (g_{\sigma N}\, \bar{\sigma})^3
- \frac{1}{4} c\, (g_{\sigma N}\, \bar{\sigma})^4
+ \frac{1}{2} m^2_\omega \, \bar{\omega}^2_0
+ \frac{1}{2} m^2_\rho \, \bar{b}^2_{03}
\nonumber \\  
&+& \frac{1}{3 \pi^2}
\sum_{N = n, p} \int^{k_N}_0 \frac{k^4}{\sqrt{k^2 + m^{*2}_N}} dk
+ \frac{1}{3 \pi^2}
\sum_l \int^{k_l}_0 \frac{k^4}{\sqrt{k^2 + m^2_l}} dk.
\label{eq:prsqhd}
\end{eqnarray}
The energy density and the pressure from the QHD-S model can be 
obtained from Eq.~(\ref{eq:eneqhd}) and Eq.~(\ref{eq:prsqhd}), respectively, 
by removing  the cubic and quartic self-interaction terms of 
the $\sigma$-meson and replacing the free meson masses with the
scaled ones given in Eq.~(\ref{eq:mscaling}).
However the pressure from the QHD-S model has
additional terms that stem from the density dependent meson masses
and coupling constants. 
Thermodynamic consistency of these terms is discussed
in Ref. \cite{smr98}, and the explicit form is given as
\begin{eqnarray} 
P &=& \frac{1}{3 \pi^2}
\sum_{N = n, p} \int^{k_N}_0 \frac{k^4}{\sqrt{k^2 + m^{*2}_N}} dk
+ \frac{1}{3 \pi^2}
\sum_l \int^{k_l}_0 \frac{k^4}{\sqrt{k^2 + m^2_l}} dk 
\nonumber \\ 
&-& \frac{1}{2} m^{*2}_\sigma\, \bar{\sigma}^2
+ \frac{1}{2} m^{*2}_\omega \, \bar{\omega}^2_0
+ \frac{1}{2} m^{*2}_\rho \, \bar{b}^2_{03} 
- m^{*2}_\sigma \bar{\sigma}^2 \frac{y}{1 + y \frac{\rho}{\rho_0}} 
\frac{\rho}{\rho_0}
- m^{*2}_\sigma \bar{\sigma} \frac{m_N}{g_{\sigma N}} 
\frac{y}{(1 + y \frac{\rho}{\rho_0})^2} \frac{\rho}{\rho_0}
\nonumber \\
&+& 
m^{* 2}_\omega \bar{\omega}^2_0
\left(\frac{y}{1 + y \frac{\rho}{\rho_0}}
- \frac{z}{1 + z \frac{\rho}{\rho_0}} \right) \frac{\rho}{\rho_0} 
+ m^{*2}_\rho \bar{b}^2_{03} 
\frac{y}{1 + y \frac{\rho}{\rho_0}}
\frac{\rho}{\rho_0}.
\label{eq:pressure-br}
\end{eqnarray}
The energy density and the pressure of the MQMC-type 
models can be obtained 
by using $g_{\sigma N} = 3 g^q_\sigma$,
$g_{\omega N} = 3 g^q_\omega$ and $g_{\rho N} = g^q_\rho$
together with proper replacement of in-medium meson masses 
in Eqs.~(\ref{eq:eneqhd}) and (\ref{eq:pressure-br}).
The resulting EoS curves for each model are shown in Fig.~\ref{fig:eos}.
%
If the pressure exerted by nuclear repulsion is strong,
the matter becomes more incompressible, which corresponds to
a large compression modulus and a stiff EoS.
The EoS's from the 
QHD, MQMC and MQMC-MB models whose $K$'s are close to
each other (310.8, 285.5 and 
324.1 MeV, respectively) show similar behaviors at the energy 
densities considered here.

The EoS from the scaling models (QHD-S and MQMC-S)
turns out to be stiffer than the EoS from other models.
This behavior can be understood by observing that
the repulsion from the $\omega-$meson 
is augmented at high densities (see Eq.~(\ref{eq:omega}) with $m_\omega$ 
replaced by $m^*_\omega$)
while the attraction caused by the $\sigma-$meson 
is not strong enough to cancel the repulsion.

\begin{figure}[tbp]
\begin{center}
\epsfig{file=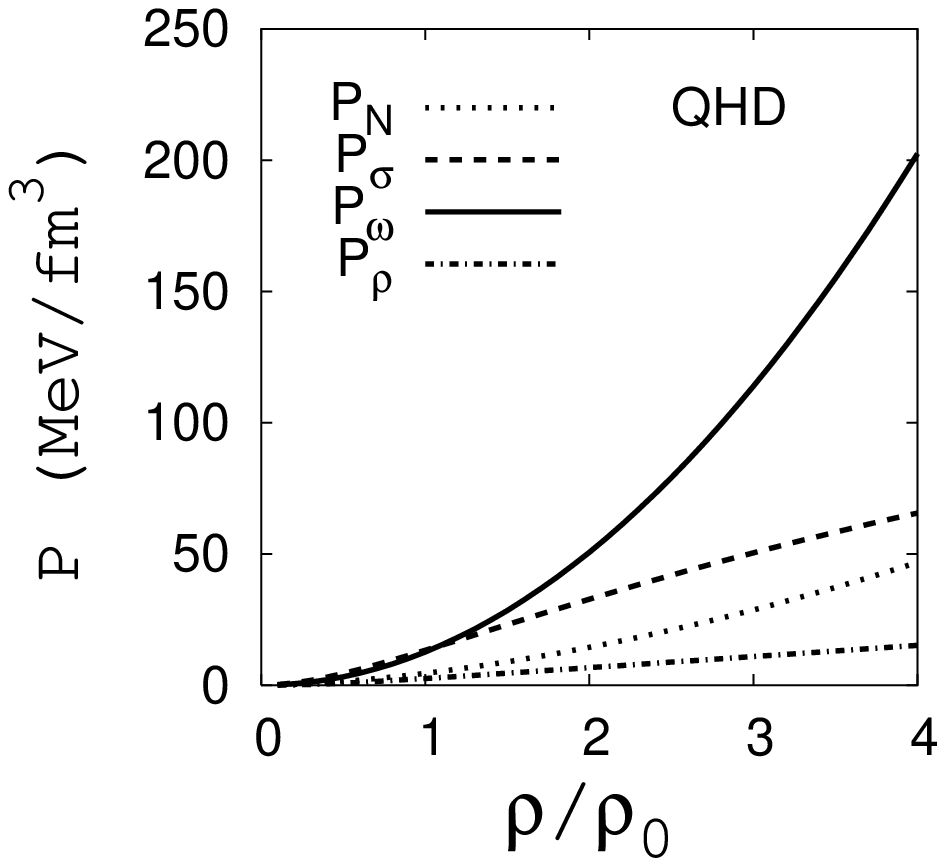,width=8.0cm}
\epsfig{file=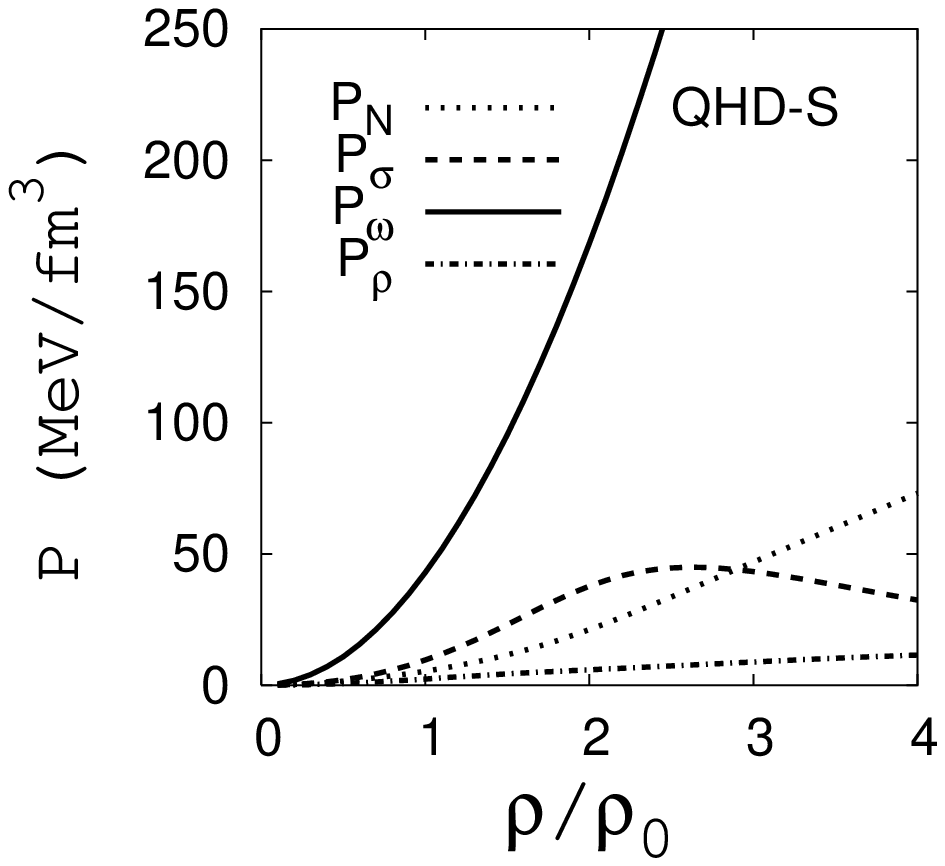,width=8.0cm}
\caption{Comparison of the magnitude of $\sigma$, $\omega$, $\rho$ and the 
nucleon contributions to the pressure for the QHD (left) and the QHD-S
(right). Definitions of $P_i$'s are given in the text.
\label{fig:mene}}
\end{center}
\end{figure}

To illustrate this argument clearly, we plot in Fig.~\ref{fig:mene}
the magnitude of the contributions to the pressure $P$ from
$N$, $\sigma$, $\omega$ and $\rho$.
Each contribution $P_i$ ($i=N,\, \sigma,\, \omega,\,\rho$) is defined as
\begin{eqnarray*}
P_N &=& \frac{1}{3 \pi^2}\sum_{N=n,p} \int^{k_N}_0
\frac{k^4}{\sqrt{k^2 + m^{*2}_N}}\, dk, \\
P_\sigma &=& \frac{1}{2} m^{*2}_\sigma\, \bar{\sigma}^2, \ \ 
P_\omega = \frac{1}{2} m^{*2}_\omega\, \bar{\omega}^2_{0}, \ \
P_\rho = \frac{1}{2} m^{*2}_\rho\, \bar{b}^2_{30}.
\end{eqnarray*}
The total pressure is approximately
\[ P \simeq -P_\sigma + P_\omega + P_\rho + P_N. \]
The remaining terms such as the cubic and quartic self-interaction
terms in QHD, extra terms for thermodynamic consistency
and lepton contributions
can be neglected since they do not determine the overall behavior of EoS.
In the case of the QHD (see the left panel of Fig. \ref{fig:mene}),  
$P_\sigma$ is non-negligible compared to $P_\omega$ in the 
density region considered here.
Since $\sigma$-meson contributes to the 
pressure negatively, $P_\sigma$ reduces the total pressure 
substantially, 
which in turn leads to a sizable softening of the EoS.
On the other hand, $P_\omega$ from QHD-S is about 3 times larger
than that of QHD, but $P_\sigma$ from QHD-S is more or less similar to
that from the QHD.
Thus the softening of the EoS due to $P_\sigma$ in QHD-S is 
relatively weak as
the density becomes high, and consequently the EoS from QHD-S is 
stiffer than that from QHD.

Here, we need to remark that in Fig. \ref{fig:eos} the EoS curves 
for MQMC-S and MQMC-MB terminate at a certain $\varepsilon$.
It is due to a breakdown of the SCC at a certain density. 
In Appendix, we show the details of why the SCC does not
have a solution at some density.

\subsection{Composition}

\begin{figure}[tbp]
\begin{center}
\epsfig{file=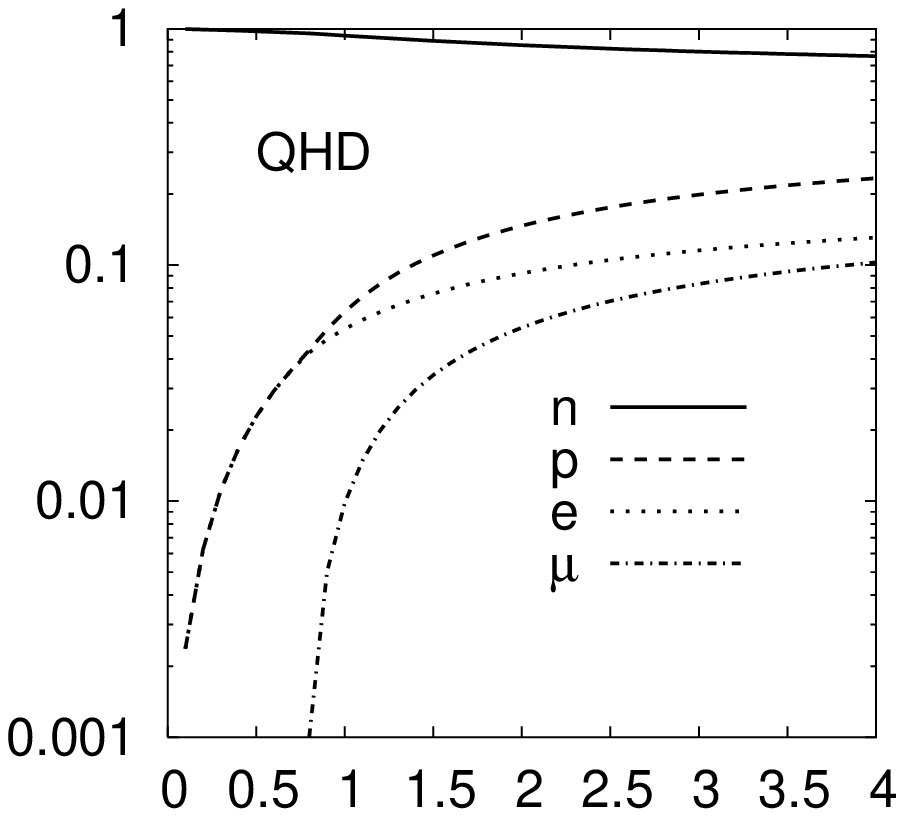,width=6.9cm}\ \ \ \ \ \ \
\epsfig{file=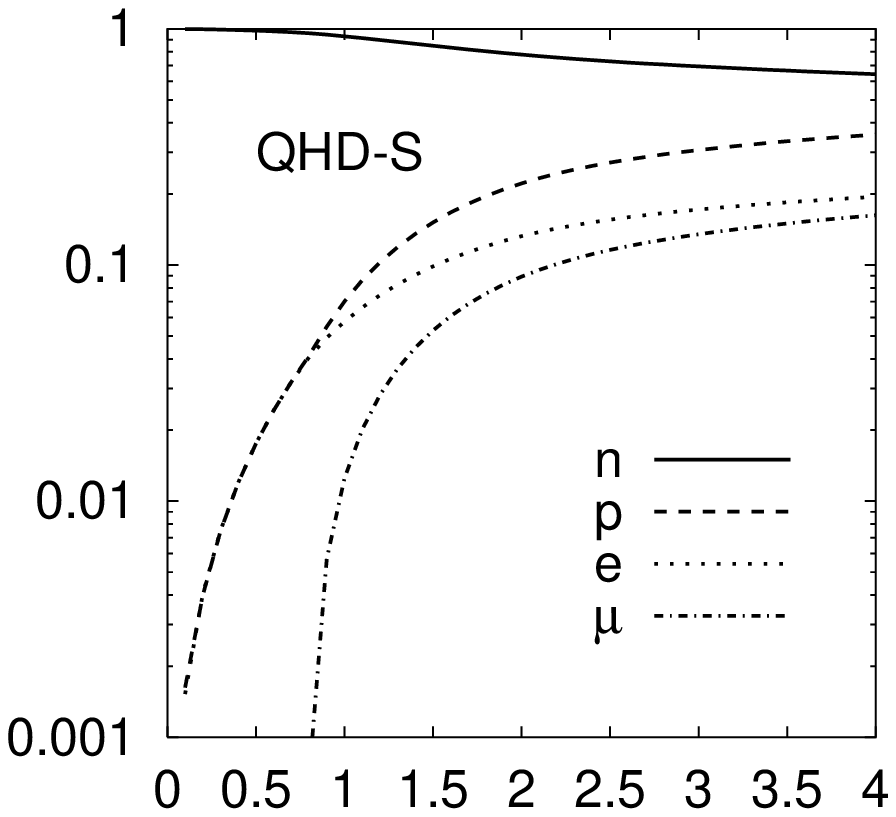,width=7.0cm} \\
\epsfig{file=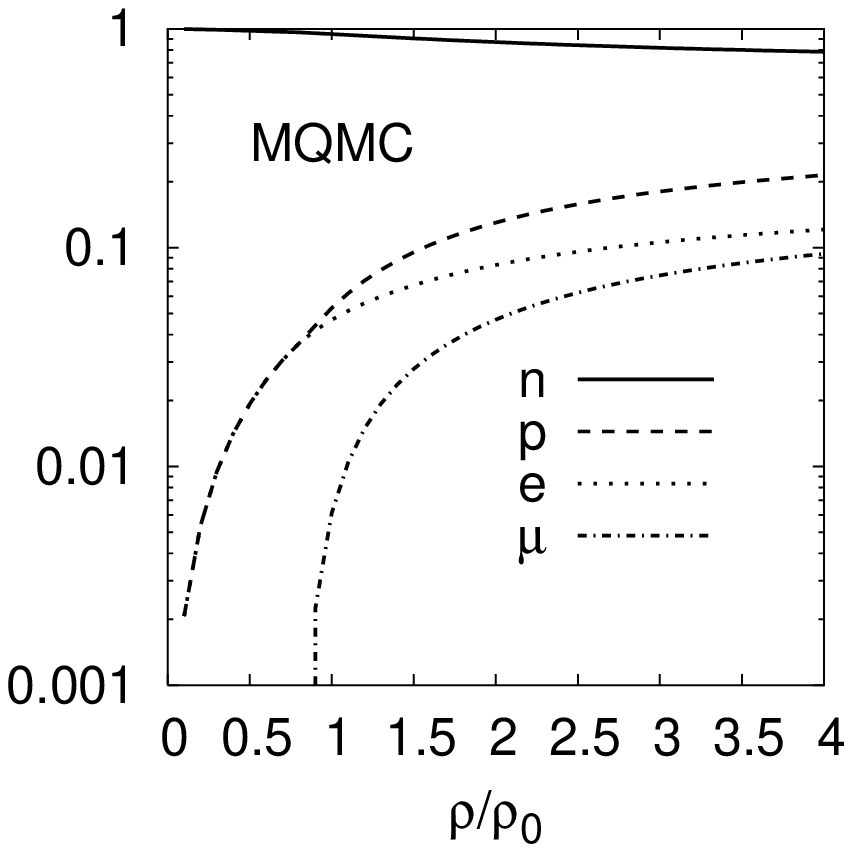,width=7.5cm}
\epsfig{file=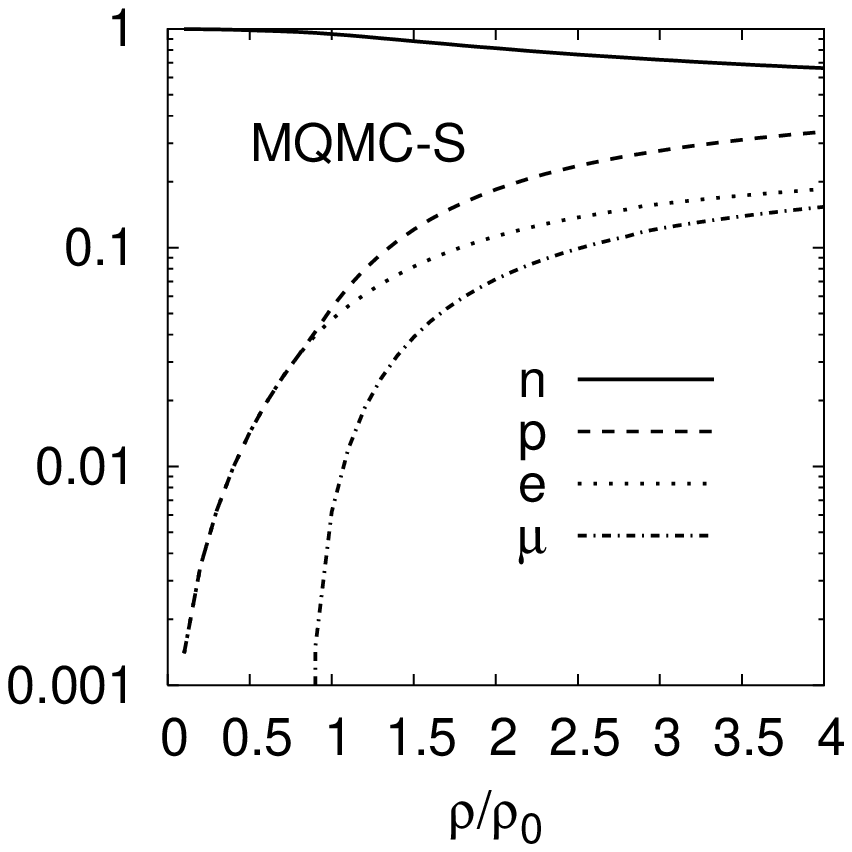,width=7.5cm}
\caption{The particle ($n$, $p$, $e$, $\mu$) fractions
in neutron stars calculated by different nuclear models are plotted against
neutron star matter densities $\rho/\rho_0$.
Since the figure for MQMC-MB is similar to that for MQMC, 
we do not show it here.
\label{fig:composition}}
\end{center}
\end{figure}

The composition of the neutron star matter is represented by the 
number of particles divided by the total baryon number.
The number of each species of particles is determined by the charge 
neutrality and the chemical equilibrium in the $\beta$-decay.
Muons can be created when the chemical equilibration condition
between the electron and the muon, $\mu_e = \mu_\mu$ can be 
fulfilled. 
Given the values of $\bar{\sigma}$, $\bar{\omega}$ and $\bar{b}_{03}$ 
for each $\rho$, charge neutrality of Eq.~(\ref{eq:charge0}),
chemical equilibrium of Eq.~(\ref{eq:chemeq}), $\mu_e = \mu_\mu$
and $\rho = \rho_n + \rho_p$ determine the number of particles
unambiguously.
The results are shown in Fig.~\ref{fig:composition}. 
%

The compositions from the QHD and MQMC models are very 
similar to each other, but the number of the protons for
the QHD-S or MQMC-S model is larger than that for the QHD or the MQMC,
respectively.
The ratio of the proton number to the neutron number,
$\rho_p/ \rho_n$ is especially important in determining the cooling
mechanism of the neutron star. 
Two major mechanisms of the cooling are emission of the
neutrinos from the interior and radiation of the photons
near the surface of the neutron star.
The simplest neutrino emission mechanism is
the so-called direct URCA (DU) process, which is nothing but
the $\beta$-decay as in Eq.~(\ref{eq:durca}), and it 
is known as the most efficient neutrino-emitting process 
in the interior of the neutron star.
In order for this process to occur, however, the ratio
$\rho_p/ \rho_n$ should be larger than a critical value.
The proton numbers from QHD-S and MQMC-S models
increase faster than those from the QHD and MQMC models 
as the density increases.
This implies that $\rho_p/ \rho_n$ from QHD-S or MQMC-S
can reach the critical 
value at lower densities and the $\beta$-decay can take place
over a wider region in the interior of the neutron star
than the QHD or MQMC models predict, causing a rapid cooling
of the neutron star.
However, a recent work on the cooling of the neutron star 
\cite{bgv-aa04} indicates that too low a threshold density
for the DU process can result in an unrealistic cooling scenario.
DU process can happen when both energy and momentum conservation
is satisfied in Eq. (\ref{eq:durca}). 
Since neutrinos have energy of the thermal fluctuation around the 
Fermi momentum of the nucleon or electron, its momentum can be
neglected. Then the momentum conservation condition for DU can be
written as $k_n = k_p + k_e$, where $k_i$ being the Fermi momentum of 
particle ``$i$". The threshold density $\rho_D$ at which
DU starts to emit neutrinos is about
$\rho_D \simeq 2.1 \rho_0$ for QHD and MQMC, and is about
$\rho_D \simeq  1.6 \rho_0$ for QHD-S and MQMC-S.
These threshold densities are too low to explain the observation 
of the neutron star temperature \cite{bgv-aa04}.
In a recent work \cite{kolo-04}, the density dependent meson masses 
and coupling constants are considered within the framework
of relativistic mean field theory.
It is shown that
they obtained lower value of $\rho_D$ than that of our QHD model.
Therefore the nuclear models we have considered in this work
may need further refinement and adjustment of parameters 
to produce threshold densities high enough for the cooling
process to be reasonable.

\subsection{Mass and radius of the neutron star}
\begin{figure}[tbp]
\begin{center}
\epsfig{file=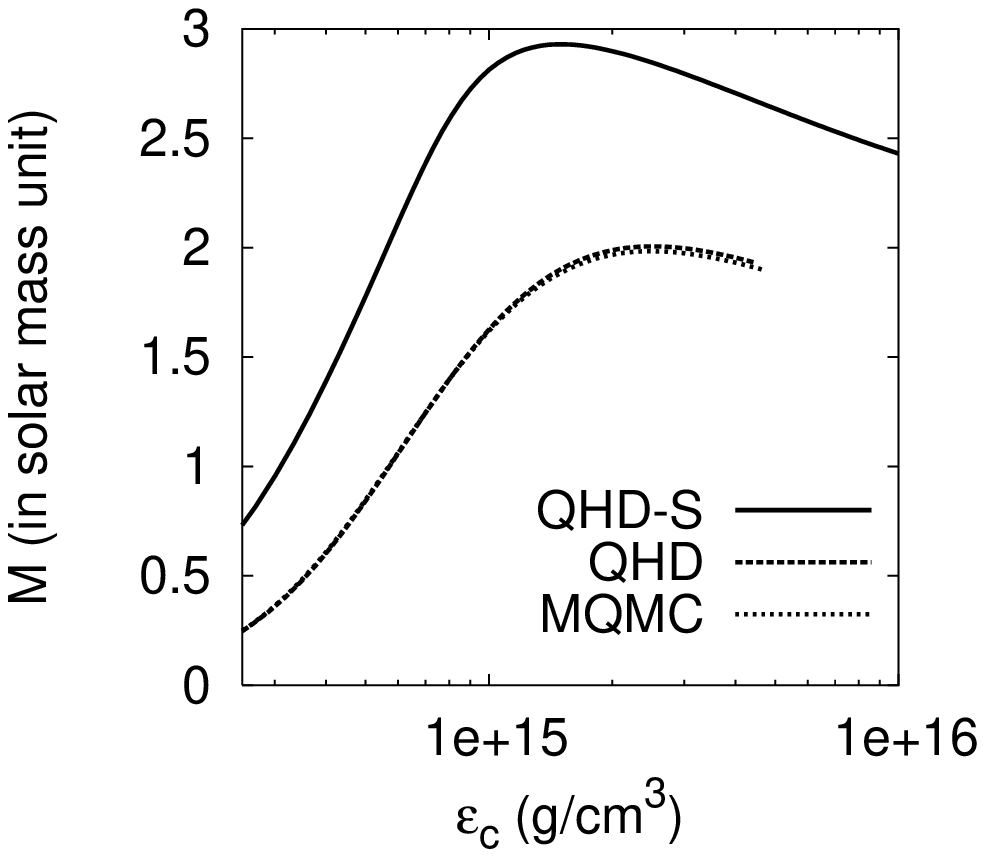, width=8.0cm}
\epsfig{file=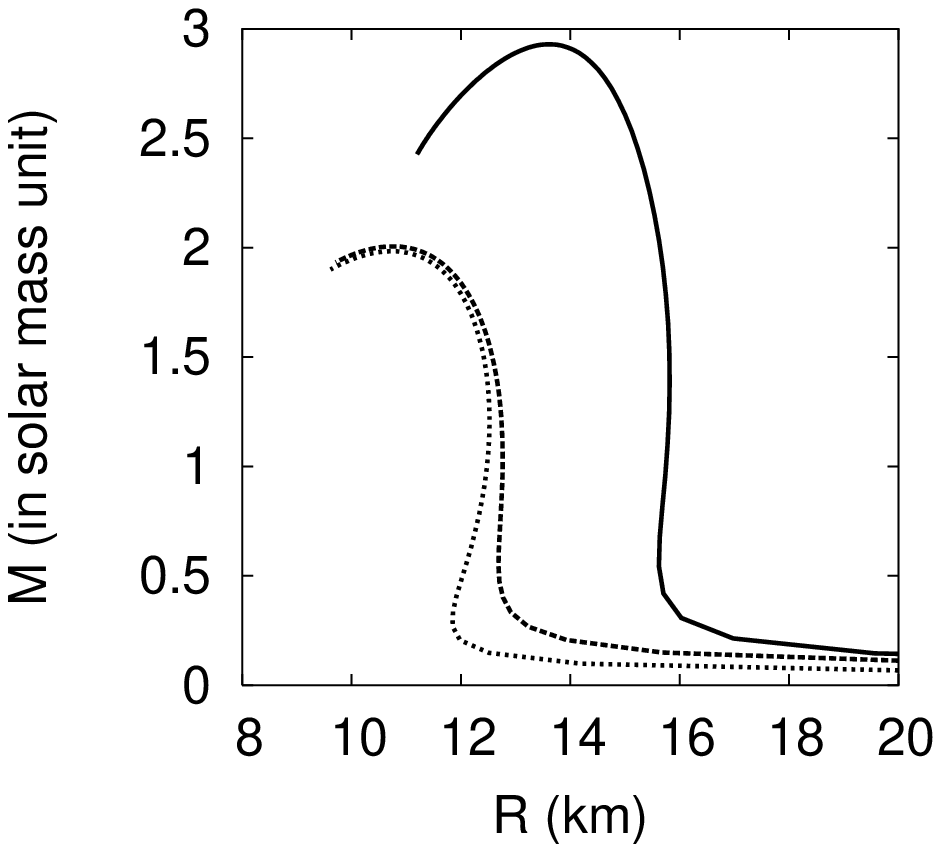, width=8.0cm}
\end{center}
\caption{Neutron star mass as a function of the central density (left) and
the radius (right). The solid line is the result from the QHD-S, 
the dashed line from the QHD, and the dotted line from the MQMC.}
\label{fig:neutronstar}
\end{figure}

The mass and the radius of the neutron star can be obtained by 
integrating Oppenheimer-Volkov equation with a given EoS.
In Fig.~\ref{fig:neutronstar}
the neutron star mass with our EoS
is plotted as a function of the central density 
$\varepsilon_c$ (left panel) and of the radius (right panel)
in units of solar mass $M_\odot$. 
The solid line is the result from the QHD-S, 
the dashed line from the QHD,
and the dotted line from the MQMC.
QHD and MQMC
give us very similar maximum mass $M_{max} \simeq 2.0 M_\odot$
with $R \simeq 10.8$ km and $\varepsilon_c \simeq 2.5 \times 10^{15}
{\rm g/cm}^3$. 
On the contrary, due to the stiff EoS the QHD-S gives us
the maximum mass of about 2.9$M_\odot$ with the radius $R \simeq 13.7$ km
and the central density 
$\varepsilon_c \simeq 1.5 \times 10^{15}{\rm g/cm}^3$.
We could not extract the maximum mass of the neutron star
for MQMC-S and MQMC-MB models 
since, as noted in the previous section, these models 
encounter a problem in solving the SCC at a density 
before reaching the maximum mass.

Recent observations of the neutron star mass from the 
radio pulsars give us the mass range of the neutron stars as
$(0.8 \sim 2.2) M_\odot$ \cite{tc98}.
The maximum mass of the QHD-S is much larger than the upper limit
of the observed values, so one may want to exclude the QHD-S among the
models for the neutron stars.
We may note, however, that more considerations are needed as follows.
First, the QHD-S produces the saturation properties successfully.
Secondly, we have considered only the nucleon degrees in this work,
but other degrees of freedom can come in such as
hyperons or the phases like meson condensation and quark matter.
In general, when these exotic degrees of freedom appear,
they reduce the Fermi momentum of nucleons
through the $\beta$-decay of the nucleons to exotic states.
This results in a smaller contribution of $P_N$ to the pressure,
and thus the EoS becomes softer with exotic degrees of freedom
than otherwise. 
Then the maximum mass of the neutron star becomes smaller.
We have confirmed this behaviour by doing
some more calculations
including hyperons in the QHD and MQMC. 
The maximum mass of $1.7 M_\odot$ is obtained for QHD,
and $1.5 M_\odot$ for MQMC. 
It will be interesting to investigate how these maximum masses
will change when medium effects on meson masses and coupling
constants are taken into consideration.
It is noted, however, that the inclusion of hyperons can give us
too low a mass limit of the neutron star \cite{bbss-prc02}, 
which is incompatible with the observations.
In Ref. \cite{bbss-prc02},
a phase transition to quark matter is considered,
and the maximum mass of the neutron star is obtained
to lie in the range $(1.4 \sim 1.7) M_\odot$.
A transition to quark matter is one possibility, 
and at the same time
transitions to other exotica are also possible.
These possibilities should be considered for a
better understanding.

\section{Summary}

In this work, we have investigated the properties of the neutron star
matter by including the effects of meson mass changes in medium.
We have explored various models at the hadron (QHD and QHD-S) 
and at the quark (MQMC, MQMC-S and MQMC-MB) levels.
We have incorporated in-medium meson masses in two ways: one by 
assuming a simple function of scaling in both QHD and MQMC models, 
and the other by treating heavy mesons as MIT bags within the 
framework of MQMC model.
The EoS and the particle fractions for the neutron
star matter are calculated.
Scaling models show appreciable difference from
the models with constant meson mass or meson bag models.
The EoS's from the scaling
models are stiffer than those from other models.
The scaling models show higher fractions of the protons than 
other models.
Stiff EoS's from the scaling models lead to heavy and large neutron stars 
whose maximum mass is about 2.9$M_\odot$ which is larger than
the observed values and the values predicted by other models 
by one solar mass.
We may, however, note that the present work is just a
first step toward the understanding of the relationship between
the in-medium properties of physical quantities (mass, coupling 
constant and etc.) and the neutron star, and 
further studies need to be done.
\section*{Acknowledgements}
This work is supported by Korea Research Foundation Grant
(KRF-2002-042-C00014).

\section*{Appendix}

In this section, we show a breakdown mechanism that occurs in 
solving the SCC of the QMC model. 
Since the SCC for MQMC-S and MQMC-MB are involved,
we employ the QMC model for the purpose of a simple illustration.
The SCC of the QMC model can be obtained by putting
$g^B_\sigma = 0$ in Eq.~(\ref{eq:sccmqmc}), which leads to
\begin{equation}
\bar{\sigma} = 3 \frac{g^q_\sigma}{m^2_\sigma}\, \rho_s\, 
C_N(\bar{\sigma}).
\label{eq:scc1}
\end{equation}
\begin{figure}[t]
\epsfig{file=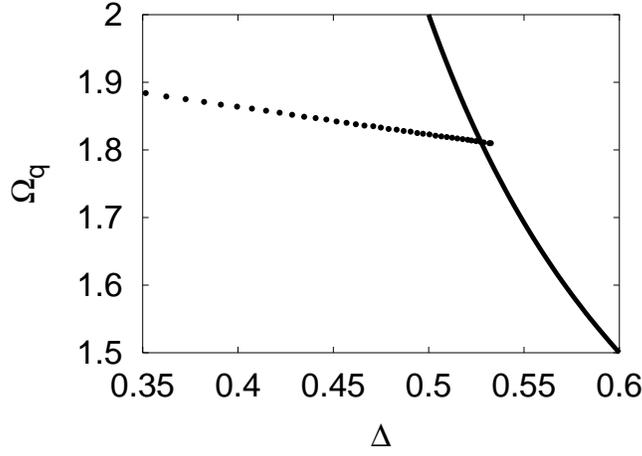, width=2.5in, angle = 270}
\caption{The solid curve represents Eq.~(\ref{eq:solution}) and the 
dots represent
$\Omega_q = \Omega_q(\Delta)$ with $R\, m^*_q$ obtained from the
QMC calculation.
In the left section divided by the solid curve, NU is greater 
than zero, which is allowed physically.
QMC calculation has been done up to $\rho = 5.3 \rho_0$.
\label{fig:Omega}}
\end{figure}
The breakdown takes place when $C_N(\bar{\sigma})$ 
changes the sign from positive to 
negative at a certain density. 
The numerator in the square bracket of Eq.~(\ref{eq:scc2}) causes
the sign change. 
Rearranging the numerator of Eq.~(\ref{eq:scc2}), we obtain
\begin{eqnarray}
{\rm NU} &=& \Omega^2_q + 3 \Omega_q(\Omega_q - 1) R m^*_q
- Z_N \left(\frac{1}{2} \Omega_q + (\Omega_q - 1) R m^*_q \right)
\label{eq:keep}\\
& & + \frac{1}{2} (R m^*_q)^2 + \frac{4}{3}\pi R^4 B 
\left(\frac{1}{2} \Omega_q + (\Omega_q - 1) R m^*_q \right),
\label{eq:neglect}
\end{eqnarray}
where NU means the numerator.
To show roughly how the sign of $C_N(\bar{\sigma})$ changes,
we make the following approximations.
First, all the terms in Eq.~(\ref{eq:neglect}) can be neglected since 
$(R m^*_q)^2$ and $R^4 B$ are very small compared to the terms 
in Eq.~(\ref{eq:keep}) in the density region we consider.
Secondly, we use $Z_N = 2$ instead of the actual value $Z_N = 2.030$.
With these approximations, we have
\begin{eqnarray}
{\rm NU} &\simeq& \Omega^2_q (3 R m^*_q + 1) 
- \Omega_q (5 R m^*_q + 1) + 2 R m^*_q \nonumber \\
&=& \left[ (3 R m^*_q + 1) \Omega_q - 2 R m^*_q \right]
(\Omega_q -1).
\label{eq:upapp}
\end{eqnarray}
Since $(\Omega_q - 1)$ is always positive,
the sign change of Eq.~(\ref{eq:upapp}) occurs at
\begin{eqnarray}
\Omega_q &=& \frac{2 R m^*_q}{3 R m^*_q + 1} \nonumber \\
&=& \frac{2 \Delta}{3 \Delta -1},
\label{eq:solution}
\end{eqnarray}
where $\Delta \equiv - R m^*_q$.
Since $\Omega_q$ is always greater than 1, the solution to 
Eq.~(\ref{eq:solution}) can exist for $\Delta > 1/3$. 
By taking $\Delta$ as an independent variable, 
we plot in Fig.~\ref{fig:Omega} $\Omega_q$
of Eq.~(\ref{eq:solution}) with the solid line.
Eq.~(\ref{eq:solution}) represents possible points 
at which the sign change occurs.
The values of $\Omega_q$ obtained from the 
numerical solution of Eq. (\ref{eq:Omega-q}) 
and the SCC of the QMC is shown 
in Fig.~\ref{fig:Omega} by the dots.
In the actual numerical calculation with the QMC model (not MQMC model),
of which we have not shown anything in this paper, 
solutions can be obtained up to $\rho = 5.3 \rho_0$ at which
$\bar{\sigma} = 34.52\ {\rm MeV}$, $R = 0.574$ and $\Delta = 0.533$.
In Fig.~\ref{fig:Omega}, 
the two curves meet at around $\Delta = 0.527$ ($\Omega_q = 1.812$),
which corresponds to the density $\rho \simeq 5.0 \rho_0$. 
If the small terms in Eq.~(\ref{eq:neglect}) are included,
the two curves will meet at around $\rho = 5.3 \rho_0$, 
which is the breakdown point of the QMC in the actual numerical 
calculation.
Similar breakdown behaviors are observed numerically 
for the MQMC-S and MQMC-MB models. 
For this reason, we could not calculate
the EoS at higher densities, and consequently 
could not extract the mass and the radius of the
neutron stars for these models.

\end{document}